# Deterministic control of all-optical analogue to electromagnetically induced transparency in coherently-coupled silicon photonic crystal cavities

Xiaodong Yang[1*], Mingbin Yu[2], Dim-Lee Kwong[2], and Chee Wei Wong[1*]

[1]*Optical Nanostructures Laboratory, Center for Integrated Science and Engineering, Solid-State Science and Engineering, and Mechanical Engineering, Columbia University, New York, NY 10027*

[2]*The Institute of Microelectronics, 11 Science Park Road, Singapore, Singapore 117685*

[*] *Correspondence and requests for materials should be addressed to either C.W.W. (e-mail: cww2104@columbia.edu) or X. Y. (e-mail: xy2103@columbia.edu).*

Quantum coherence in atomic systems has led to fascinating outcomes, such as laser cooling and trapping, Bose-Einstein condensates, and electromagnetically-induced-transparency (EIT). In EIT, the sharp cancellation of medium absorption[1] has led to phenomena such as lasing without inversion[2], freezing light[3], and dynamic storage of light in a solid-state system[4]. Similar to atomic systems, EIT-like effects can be observed through classical and optical means. Here we report the first experimental deterministic tuning of all-optical analogue to EIT in coherently-coupled standing-wave photonic crystal cavities. Our observations include transparency-resonance lifetimes more than three times the single loaded cavity,

**Fano-type lineshapes, and stepwise control of the coherent cavity-cavity interference. Our system, with wavelength-scale localization and coupled to a single waveguide, is analyzed well through the coupled-mode formalism which examines the delay in both transparency- and Fano-like lineshapes. Our observations support applications towards all-optical trapping, stopping and time-reversal of light in a solid-state scalable implementation.**

Coherent interference between normal modes in coupled optical resonators can induce a sharp transparency window in an otherwise non-transmitting background, affording an additional degree of freedom in controlling dispersion. Several recent theoretical proposals have suggested this intriguing possibility in an optical analogy to EIT[5-10], towards stopping light[6] and trapping light beyond the fundamental delay-bandwidth product[11] at room temperature. Slow-light in photonic structures have been examined[12-16] but is likewise bounded by the fundamental delay-bandwidth limit. Recently EIT-like effects were examined experimentally in coupled whispering-gallery mode resonators, with observations of slow group velocities and storing light on-chip beyond the static delay-bandwidth limit[17-20]. These are traveling-wave resonators with modal volumes of tens of cubic wavelengths ($\sim 10(\lambda/n)^3$ or more). Here, we present the first experimental observation of an all-optical analogue to EIT in defect-type standing-wave photonic crystal cavities, with wavelength-scale photon localization. Our cavities have an order of magnitude stronger localization ($\sim(\lambda/n)^3$), are fundamentally single-mode (in contrast to forward- and backward-scattering in whispering-gallery resonators), and require only a single waveguide for coherent interference. These observations are enabled by precise nanofabrication, and deterministic phase- and resonance-matching measurements. Our coupled photonic crystal cavities are integrated on-chip without the need for atomic resonance, removing much of the limitations on bandwidth and decoherence in atomic systems, for applications in optical communications.

Our EIT-like system, depicted in Fig. 1, consists of a photonic crystal waveguide side-coupled to two high intrinsic quality factor (*Q*) photonic crystal cavities, such as examined in Refs. [6, 21]. Such cavities allow wavelength-scale photon localization with ultrahigh-*Q* factors[22, 23], for nonlinear[24, 25] and quantum optics[26]. Our particular cavity examined is a defect-type cavity, formed with three missing air holes (*L3*) in an air-bridge triangular-lattice photonic crystal membrane. The membrane has a thickness of 0.6*a* and hole radius of 0.29*a*, where the lattice period *a* = 420 nm. In each cavity, the nearest neighbour holes at the cavity edge are shifted ($s_1$) by 0.15*a* to tune the radiation mode field for increasing the intrinsic *Q* factors[27]. The center-to-center waveguide-to-cavity separation is $2\sqrt{3}a$, and the in-plane separation *L* between the two cavities is 11*a* and 10*a* for sample 1 and sample 2, respectively. We optimized the cavity *Q*s with $s_1$ tuning of the nearest neighbour holes, computed directly from full three-dimensional (3D) finite-difference time-domain (FDTD) numerical simulations[28]. For a single tuned ($s_1$ = 0.15*a*) *L3* cavity coupled to the photonic crystal waveguide with $2\sqrt{3}a$ lattice spacing, we designed the intrinsic *Q* factor $Q_{int}$ as ~ 60,000, the waveguide-cavity coupling *Q* factor $Q_c$ ~ 1,600, the total *Q* factor $Q_{tot}$ ~ 1560, and the modal volume *V* is ~ 0.74 cubic wavelengths [$(\lambda/n)^3$]. In this work, we deliberately designed the ratio of $Q_{int}$ to $Q_c$ to be high (~ 37.5) in each cavity, so as to operate in the overcoupled regime with (vertical) radiation loss well-suppressed for in-plane cavity-cavity interference. With tuning of three air holes at cavity edge, the $Q_{int}$ can also reach up to 100,000, further reducing radiation losses[29]. When the two cavity resonances, $\omega_1$ and $\omega_2$, overlap (with $\delta = 2\tau_{total,1}(\omega_1-\omega_2) \neq 0$, where $\tau_{total,1}$ is the loaded photon lifetime of cavity 1) and the cavity-cavity round-trip phase $2\phi$ satisfy the condition of forming a Fabry-Pérot resonance ($2n\pi$), the current system represents an all-optical analogue of EIT, with two nonorthogonal modes[30] at the frequency of $(\omega_1+\omega_2)/2$. When $\delta < \sim 3.5$, the linewidth of the transparency peak is narrower than each individual loaded cavity resonances, achieving EIT-like coherent interference. Fig. 1(c) shows the *E*-field intensities of transparency mode of the two coupled *L3* cavities. One mode decays slower compared to a single cavity, the other mode decays faster. The mode with the slower decay is responsible for the EIT-like spectral feature[30].

The devices are fabricated in a silicon-on-insulator substrate, with 248 nm UV lithography in a CMOS foundry for advanced integrated circuits. The scanning electron micrograph (SEM) in Fig. 1(b) shows remarkably uniform photonic crystal structures. The fabrication disorder is statistically parameterized (see Methods) with resulting lattice period of 422.97 ± 1.65 nm, hole radius of 121.34 ± 1.56 nm, feature ellipticity of 1.57 ± 0.79 nm, remaining edge roughness of 1.66 nm, and edge roughness correlation length of 18 nm [31]. The single-crystal device layer is 250 nm thick and on top of a 1 μm buried oxide; the buried oxide is subsequently sacrificially etched to form air-bridged membrane structures such as shown in Fig. 1(b). An in-line fiber polarizer with a polarization controller is used to couple transverse-electric polarization light from an amplified spontaneous emissions source (ranging from 1525 nm to 1610 nm) into the waveguide via a tapered lensed fiber. A second tapered lensed fiber collects the transmission from the waveguide output, and the signal sent to an optical spectrum analyzer with 10-pm resolution.

In the design, we implemented cavities with the same resonance; due to actual fabrication mismatches, there is typically a few-nm wavelength difference between the two cavities in our fabricated samples (over 50+ devices measured in several different chips). The cavity-cavity round-trip phase $2\phi$ also needs to be actively tuned to match the $2n\pi$ condition at particular frequencies. To align the cavity resonances, a digital tuning method recently demonstrated can be employed[31]; instead here we use a thermo-optic method to tune both $\delta$ and $\phi$ simultaneously. A 532 nm (above silicon band gap) continuous-wave laser is focused to within a 5 μm spot size at one cavity region through an objective lens to locally and thermo-optically perturb the silicon refractive index (with a *dn/dT* of $1.85\times10^{-4}$/K at room temperature). The pump position and the pump power are carefully selected so that both $\delta$ and $\phi$ are precisely controlled for coherent interference.

We present here two series of experiments: the first (Fig. 2) with the coupled cavities initially mismatched, and the second (Fig. 3) with the coupled cavities initially phase- and resonances-matched for EIT-like spectrum without any external pumping. In Fig. 2(a), the black, solid curves illustrate, for the first time, the measured transmission spectra of the EIT-like photonic crystal system for various pump powers (from 0 to 1.40 mW) for sample 1. Fig. 2(a)(i) shows the initial transmission spectrum for two uncoupled standing-wave cavities, with two separated Lorentzian lineshapes, where the cavity resonances are $\lambda_1$ = 1548.63 nm and $\lambda_2$ = 1549.45 nm respectively (differing due to fabrication mismatch). As the pump power is increased, both cavity resonances are shifted to longer wavelengths and the resonance detuning between two cavities ($\delta$) is deterministically narrower. For example, when the pump power is 0.28 mW, the detuning $\delta$ is 2.74, and an EIT-like transparency peak distinctively appears. With further decrease in detuning $\delta$, Fig. 2(a)(ii)-(iv), the transparency peak get progressively narrower, supporting increased interference between the two cavity modes. The full width at half maximum (FWHM) of the symmetric transparency peak in Fig. 2(a)(iv) is ~ 0.12 nm, or a $Q_{EIT}$ of 13,000, sizably larger than the loaded $Q$ of *each L3* cavity and a longer photon delay than both non-interacting cavity lifetimes combined. For the single cavity, the measured total (loaded) quality factors are $Q_{tot,1}$ ~ 4,000 and $Q_{tot,2}$ ~ 3,600, estimated from the FWHM. The discrepancy between calculated and measured $Q$s is due to fluctuation of the resonance frequency of the cavity mode[32]. In the differential pumping of the two cavities, we located the pump spot closer to cavity 1 and hence it has a larger frequency shift than cavity 2. Moreover, we note that under stronger pump excitation, Fig. 2(a)(v)-(vi) now show Fano-like lineshapes, with the highly asymmetric (sharp on/off transmission) edge shifting from the higher-frequency edge (Fig. 2(a)(v)) to the lower-frequency edge (Fig. 2(a)(vi)), due to departure of the round-trip phase ($2\phi$) from $2n\pi$. Furthermore, we note the limit in the minimum linewidth of the transparency peak is not due to disorder scattering, but coupling into the overall EIT-like system. The wavelength shift per mW pump power is linear and ~ 1.32 nm/mW. We also show only the raw (unfiltered) measurement data, where the Fabry-Perot noise oscillations away from the central resonances are due to waveguide facet reflections, which can be removed with an integrated fiber-to-strip waveguide spot-size converter.

These EIT-like observations in wavelength-scale coupled photonic crystal cavities are further confirmed with sample 2 as shown in Fig. 3(a). Here we now have two cavities with initial small resonance detuning (~ the single cavity linewidth), and the resonance wavelengths are $\lambda_1$ = 1549.10 nm and $\lambda_2$ = 1548.80 nm, respectively. The measured total quality factors are $Q_{tot,1}$ ~ 4,100 and $Q_{tot,2}$ ~ 3,100. Without any external pumping, the transmission spectrum already shows asymmetric Fano-like lineshapes [in Fig. 3(a)(vi)]. Upon pump tuning, the two cavity resonances *separate* and the strong in-plane coherent coupling results in clear observations of the EIT-like lineshapes, as shown in Fig. 3 (a)(ii)-(v). The FWHM of the transparency peak in Fig. 3(a)(v) is ~ 0.15 nm, or a $Q_{EIT}$ of 10,400. Furthermore, we verified the mode field distributions under controlled tuning. Fig. 4 shows examples of near-field radiation patterns from the defect cavities, measured with an infrared camera with an incident narrow-linewidth tunable laser diode. At the resonance wavelength of each *L3* cavity, the radiation exhibits a single bright spot (Fig. 4(a) and 4(c)) at the spatial location of each cavity. In contrast, exactly at the EIT-like transparency frequency, the mode field distribution shows radiation from both cavities (Fig. 4(b)), further confirming coherent interference between the two coupled cavities. Future experiments also include samples with larger physical in-plane separation (*L*) between the two cavities, so that the cavity-cavity detuning and phase difference can be pumped separately with two beams.

The measured transmission spectra are next examined with the coupled-mode formalism[33], with theoretical model shown in Fig. 1(a). The dynamical equation for the two side-coupled cavity mode amplitudes are[34]

$$\frac{da_1}{dt} = \left(-\frac{1}{2\tau_{total,1}} + i(\omega_1 + \Delta\omega_1 - \omega_{wg})\right)a_1 + \kappa s_{1+} + \kappa s_L \quad (1)$$

$$\frac{da_2}{dt} = \left(-\frac{1}{2\tau_{total,2}} + i(\omega_2 + \Delta\omega_2 - \omega_{wg})\right)a_2 + \kappa s_R + \kappa s_{2+} \quad (2)$$

where $a$ is the cavity mode amplitude normalized to represent the cavity mode energy $U$ ($= |a|^2$) and $s$ the waveguide mode amplitude normalized to represent the power of the waveguide mode $P = |s|^2$. In addition, as shown in Fig. 1(a), $s_{1-} = \exp(-i\phi)s_L + \kappa a_1$, $s_R = \exp(-i\phi)s_{1+} + \kappa a_1$, $s_L = \exp(-i\phi)s_{2+} + \kappa a_2$, and $s_{2-} = \exp(-i\phi)s_R + \kappa a_2$. $\phi = \omega_{wg} n_{eff} L / c$ is the phase difference between the two cavities, where the effective index of fundamental mode in photonic crystal waveguide $n_{eff}$ is 2.768 at 1.55 μm. $\kappa$ is the coupling coefficient between the waveguide mode $s(t)$ and the cavity resonance mode $a(t)$, and $\kappa = i\exp(-i\phi/2)/\sqrt{2\tau_c}$. In Eq. (1) and Eq. (2), without considering the nonlinear absorption terms, the total loss rate for the resonance mode $1/\tau_{total}$ [Ref. 25] is $1/\tau_{total} = 1/\tau_c + 1/\tau_{int}$, where $1/\tau_c (= \omega/Q_c)$ and $1/\tau_{int} (= \omega/Q_{int})$ are the decay rates from the cavity into the waveguide and into the continuum respectively. The transmission coefficient $T = |s_{2-}|^2 / |s_{1+}|^2$ is solved numerically through Equations (1) and (2) (see Methods).

The red, dashed curves in Fig. 2(a) and Fig. 3(a) show the theoretical results for both samples with remarkable fits. When the round-trip phase $2\phi$ is close to $2n\pi$, the transmission spectrum exhibits a narrow and almost symmetric EIT-like peak, as shown in Fig. 2(a)(ii)-(iv) and Fig. 3(a)(ii)-(v). As the phase difference shifts away from $2n\pi$, the asymmetric Fano-like lineshape is observed at one side of the transmission dip. Fig. 2(b) and Fig. 3(b) show the calculated corresponding phase shift in transmission for sample 1 and sample 2. At the EIT-like transparency region, the phase slope is positive and shows a steep linear normal dispersion. As the coupling strength between two cavities is increased and tuned by the pump beam, the slope is steeper, indicating a longer photon delay. For EIT-like lineshapes, the optical delay is limited by the resonance detuning between two cavities. The red dots show the locations of EIT-like peaks or Fano-like transitions where slow light is expected, in addition to the fast light regions at the two cavity resonances. Fig. 5(a) and Fig. 5(b) show the calculated corresponding optical delays at the EIT-like peaks in Fig. 2(a)(iv) and Fig. 3 (a)(v), with a temporal delay of 20 ps and 16 ps respectively. We note that even with the measured Fano-like transition in Fig. 3 (a)(vi), a 12.5 ps optical delay can be observed (Fig. 5(c)) when tuned to the Fano-like edge, with a 5 ps pulse advance when detuned within 41 pm from the Fano-like edge in the measured transmission of our coupled photonic crystal cavities.

In summary, we achieved for the first time an all-optical analogue to EIT in coherently-coupled wavelength-scale photonic crystal cavities, through deterministic control of the resonance detuning and cavity-cavity phase-matching. In our standing-wave cavities with $(\lambda/n)^3$ localization, distinctive EIT- and Fano-like lineshapes are observed in the coupled cavity interactions, with measured EIT-like linewidths narrower than individual resonances. The maximum $Q$ of our measured transparency-resonance is 13,000, corresponding to an optical delay up to 20 ps. Our experimental and theoretical results support efforts towards realization of photon pulse trapping, dynamic bandwidth compression, and nonlinear optics in integrated low-power photonic crystal cavity arrays.

**Acknowledgements**

This work was partially supported by the New York State Office of Science, Technology and Academic Research. Xiaodong Yang acknowledges the support of an Intel Fellowship.

Correspondence and requests for materials should be addressed to C. W. W and X. Y.

**References**1. S. E. Harris, Electromagnetically induced transparency, *Physics Today* **50**, 36 (1997); M. D. Lukin and A. Imamoglu, Controlling photons using electromagnetically induced transparency, *Nature* **413**, 273 (2001).

2. S. E. Harris, Lasers without inversion: Interference of lifetime-broadened resonances, *Phys. Rev. Lett.* **62**, 1033 (1989); A. Imamoglu and S. E. Harris, Lasers without inversion: interference of dressed lifetime-broadened states, *Opt. Lett.* **14**, 1344 (1989).

3. L. V. Hau, S. E. Harris, Z. Dutton, and C. H. Behroozi, Light speed reduction to 17 metres per second in an ultracold atomic gas, *Nature* **397**, 594 (1999); C. Liu, Z. Dutton, C. H. Behroozi, and L. V. Hau, Observation of coherent optical information storage in an atomic medium using halted light pulses, *Nature* **409**, 490 (2001); D. F. Phillips, A. Fleischhauer, A. Mair, R. L. Walsworth, and M. D. Lukin, Storage of light in atomic vapor, *Phys. Rev. Lett.* **86**, 783 (2001); O. Kocharovskaya, Y. Rostovtsev, and M. O. Scully, Stopping light via hot atoms, *Phys. Rev. Lett.* **86**, 628 (2001).

4. J. J. Longdell, E. Fraval, M. J. Sellars, and N. B. Manson, Stopped light with storage times greater than one second using electromagnetically induced transparency in a solid, *Phys. Rev. Lett.* **95**, 063601 (2005).

5. D. D. Smith, H. Chang, K. A. Fuller, A. T. Rosenberger, and R. W. Boyd, Coupled-resonator-induced transparency, *Phys. Rev. A* **69**, 063804 (2004).
10

Figure Legends

**Figure 1 | Designed and fabricated EIT-like photonic crystal system. a,** Schematic of EIT-like system including a waveguide side-coupled to two cavities. **b,** SEM of two photonic crystal *L3* cavities side-coupled to a photonic crystal waveguide. Each cavity has the nearest neighbour holes at the cavity edge tuned by $s_1 = 0.15a$ to suppress the vertical radiation loss by more than an order of magnitude compared to the coupling rate. Background: SEM of cavity array under fabricated for measurements. **c,** $E_y$-field of the coupled-cavity transparency mode at mid-slab. Inset: *k*-space amplitudes for single *L3* cavity, illustrating high radiation *Q*.

**Figure 2 | Measured and theoretical transmission lineshapes and phase shift for sample 1. a,** Measured and theoretical transmission lineshapes with various detuning ($\delta = 2\tau_{total,1}(\omega_1-\omega_2)$ where $\tau_{total,1}$ is the loaded lifetime of cavity 1) with initial *large* cavity-cavity detuning (sample 1; $L = 11a$). Solid black lines show experimental data and red dashed lines show theoretical fits. **b,** Corresponding theoretical transmission phase shift. The red dots show the locations of EIT-like peaks or Fano-like transitions where slow light is expected.

**Figure 3 | Measured and theoretical transmission lineshapes and phase shift for sample 2. a,** Measured and theoretical transmission lineshapes with various detuning, with initial *small* cavity-cavity detuning (sample 2; $L = 10a$). Solid black lines show experimental data and red dashed lines show theoretical fits. **b,** Corresponding theoretical

transmission phase shift. The red dots show the locations of EIT-like peaks or Fano-like transitions where slow light is expected.

**Figure 4 | Near-field images of the light emitted from the defect region under 0.32 mW pump power for sample 1. a** and **c,** Off the EIT transparency peak, at the cavity resonances ($\lambda_1$ = 1549.08 nm and $\lambda_2$ = 1549.55 nm for cavity 1 and cavity 2, respectively), only a single cavity predominantly radiates. **b,** Both cavities radiate exactly at the EIT transparency peak ($\lambda_{EIT}$ = 1549.29 nm) observed earlier. The superimposed grey dotted lines depict the same position of the photonic crystal waveguide, and are guides to the eye.

**Figure 5 | Corresponding calculated optical delays for the coherent interference measurements.** Panel **a** and **b** correspond to the EIT-like lineshapes measured in Fig. 2(a)(iv) and Fig. 3(a)(v) respectively. Panel **c** correspond to the Fano-like lineshapes measured in Fig. 3(a)(vi).

**Methods**

**Fabrication**

The photonic crystal is designed and fabricated as a hexagonal lattice of air-holes arranged in a silicon slab ($n = 3.48$) with 250 nm thickness on a 1 μm buried oxide. The photonic crystals has a lattice constant $a$ of 420 nm and a hole radius $r$ of $0.29a$. The 248 nm UV lithography in a CMOS foundry at the Institute of Microelectronics has 180 nm critical dimension resolution. The buried oxide is subsequently sacrificially removed to create a suspended membrane, and cleaved for our measurements. The scanning electron micrograph (SEM) in Fig. 1(b) shows remarkably uniform photonic crystal structures.

Statistical parameterization[35] shows good uniformity in the structure, analyzed with the methods described in Skorobogatiy *et al.*. The edge detection algorithm used to do the image disorder quantification, involves categorizing the image into holes and the substrate region. First, we normalize the image pixel to be distributed between 0 and 1. Then, each pixel of the image is compared to an optimum threshold parameter, which is chosen based on the histogram of the pixel value of the image. For our scanning electron microscope image (resolution of 1.10 nm), a threshold value of 0.28 was chosen. The average radius was found to be 121.34 ± 1.56 nm and a RMS deviation of an edge from such a circle was calculated to be 2.08 nm. The lattice period was found to be 422.97 ± 1.65 nm. A statistical analysis of the hole ellipticity was also carried out to give a feature ellipticity of 1.57 ± 0.79 nm, the direction of an ellipse major axis of 0.60 ± 17.06°, and a RMS deviation of an edge from such an ellipse of 1.66 nm. Since the error in the parameter for the direction of an ellipse major axis is larger than the estimated parameter value we conclude that ellipticity is not really a statistically significant variation in our features, but rather a part of the edge roughness. To study the roughness of features in our PhC lattice a fractal methodology as proposed by Skorobogatiy *et al.*[35] has been employed. A correlation length of 18 nm was calculated using the parameterization of the "height-to-height" correlation function as proposed in Skorobogatiy *et al.*.

**Theoretical analysis**

Equations (1) and (2) describe the dynamic behavior of coherent interference between two coupled photonic crystal cavities, and is numerically integrated with a variable order Adams-Bashforth-Moulton predictor-corrector method (*Matlab*® ode113 solver)[25]. All the parameters used in calculation are from either FDTD simulation or experiments. The resonance of each cavity is obtained from the minima of the transmission spectra. $Q_{int}$ is 60,000 from our 3D FDTD simulations. $Q_c$ of each cavity is slightly adjusted around the measured values to fit the background shape of the transmission dip. The phase difference $\phi$ is tuned to obtain the similar transmission lineshapes as experimental data.

To support the coupled-mode modeling, we utilize FDTD numerical simulations with a rectangular computational domain, $L_x \times L_y \times L_z$, of dimensions $10.5 \times 6.3 \times 1.68$ μm³, surrounded by a 0.42 μm-thick perfectly matched layer (PML). The reflectivity of the PML was set to $10^{-9}$, so that the wave reflections at the boundaries are negligible. A $21 \times 21 \times 21$ nm³ uniform grid covered both the computational domain and the PML. The quality factor $Q$ is directly calculated from the total energy in the cavity and the power flow from the cavity.

**Figure 1**

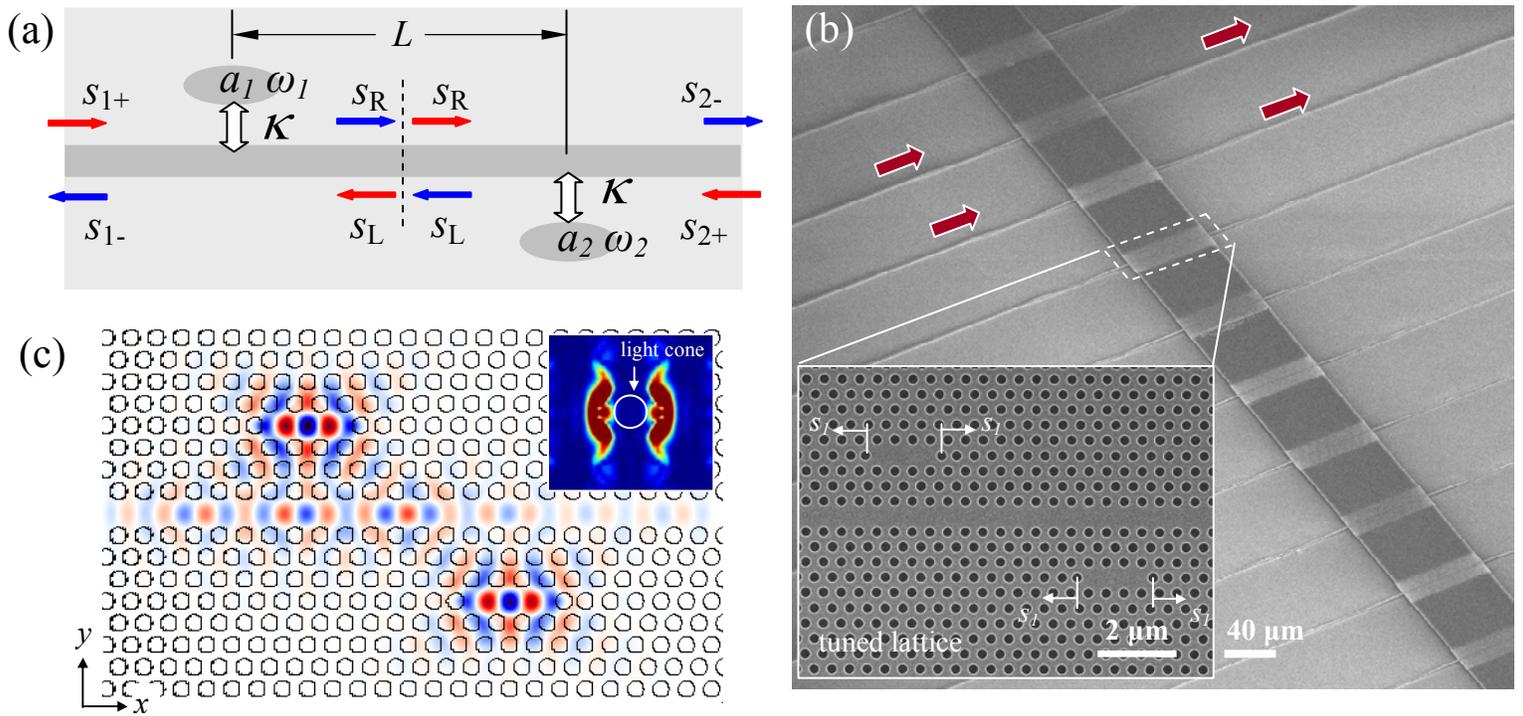

**Figure 2**

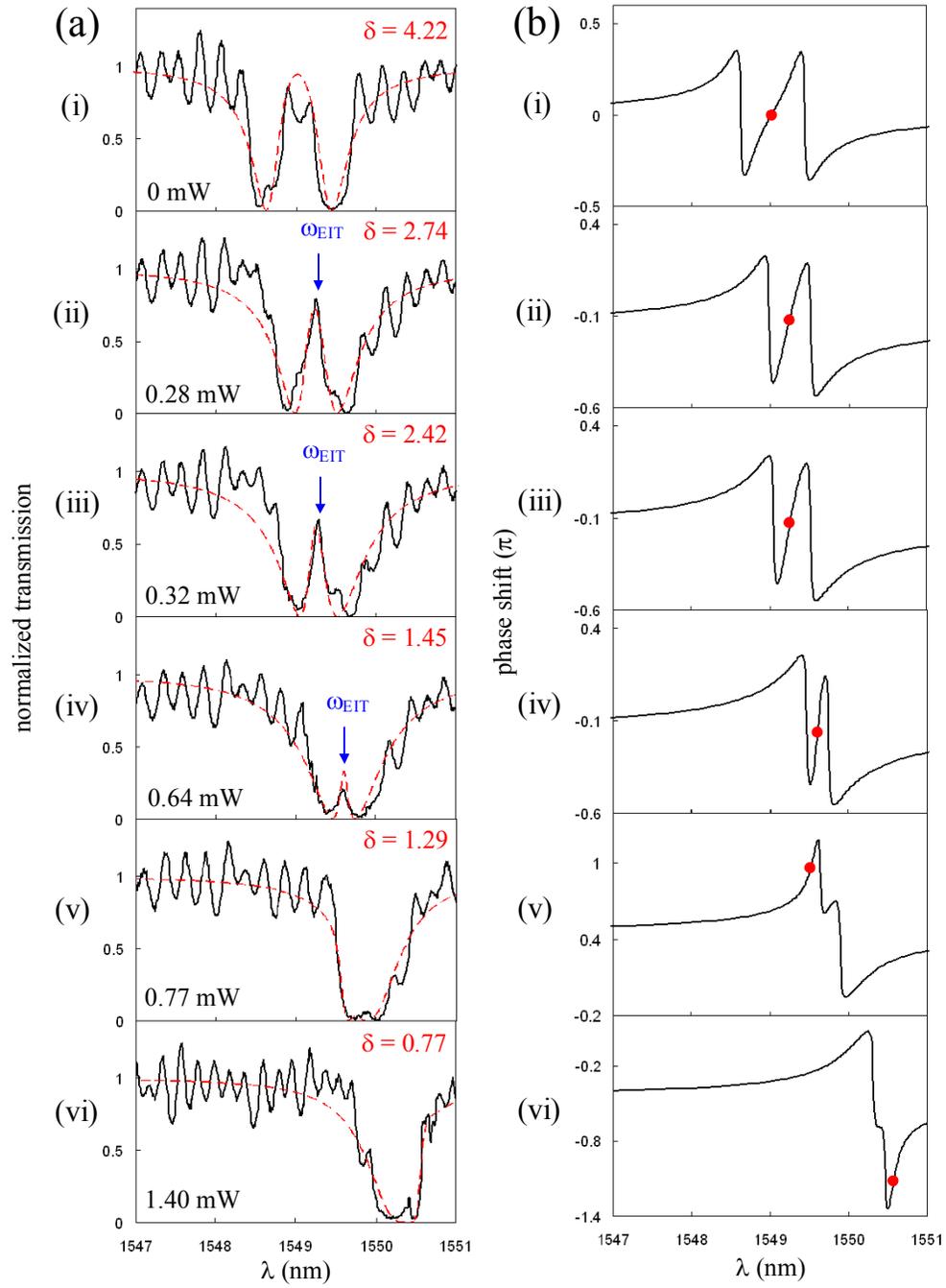

**Figure 3**

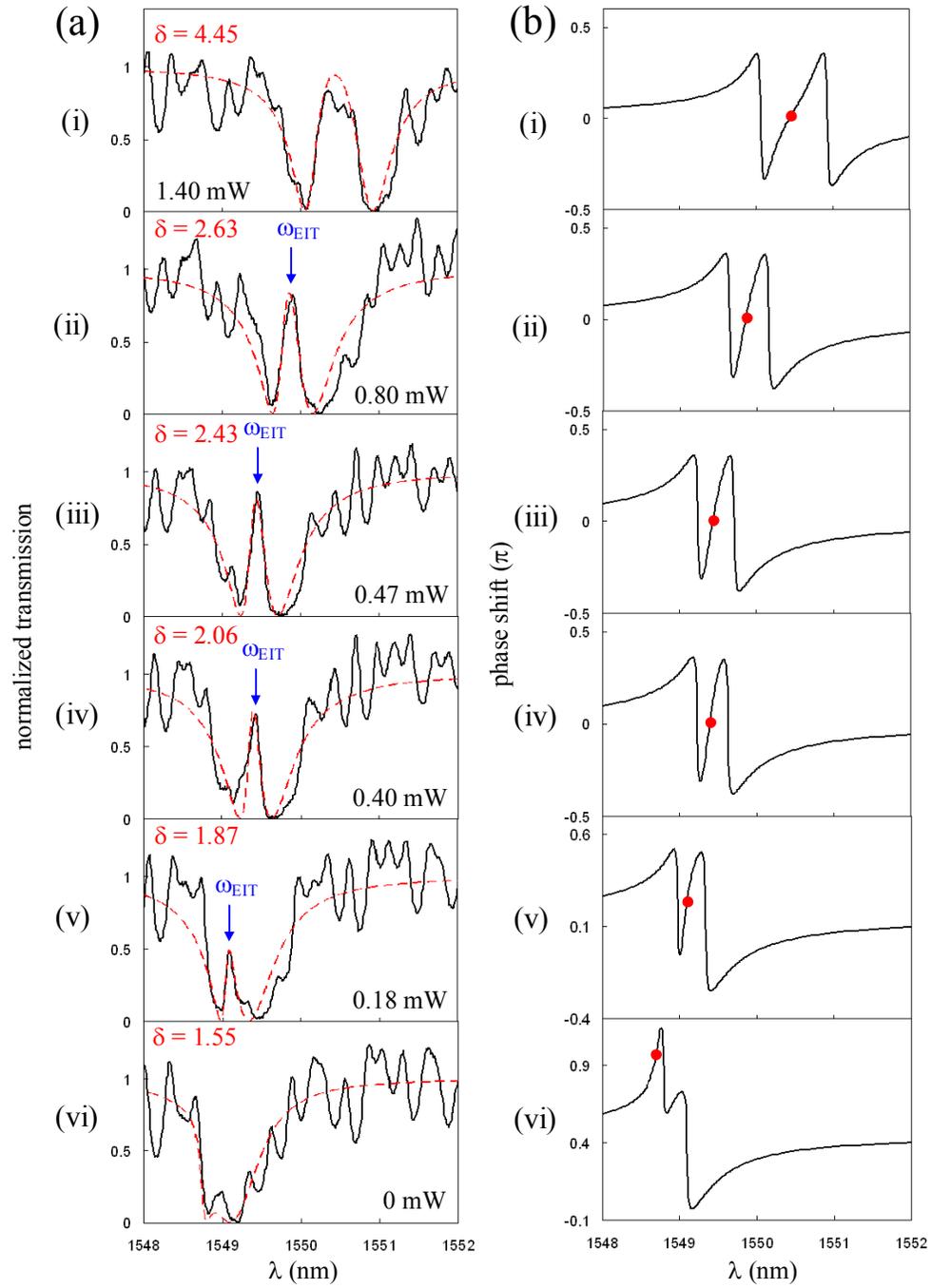

**Figure 4**

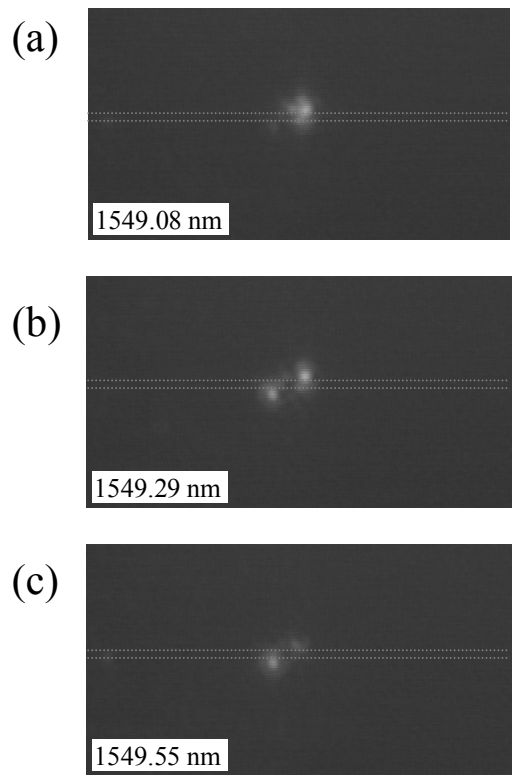

**Figure 5**

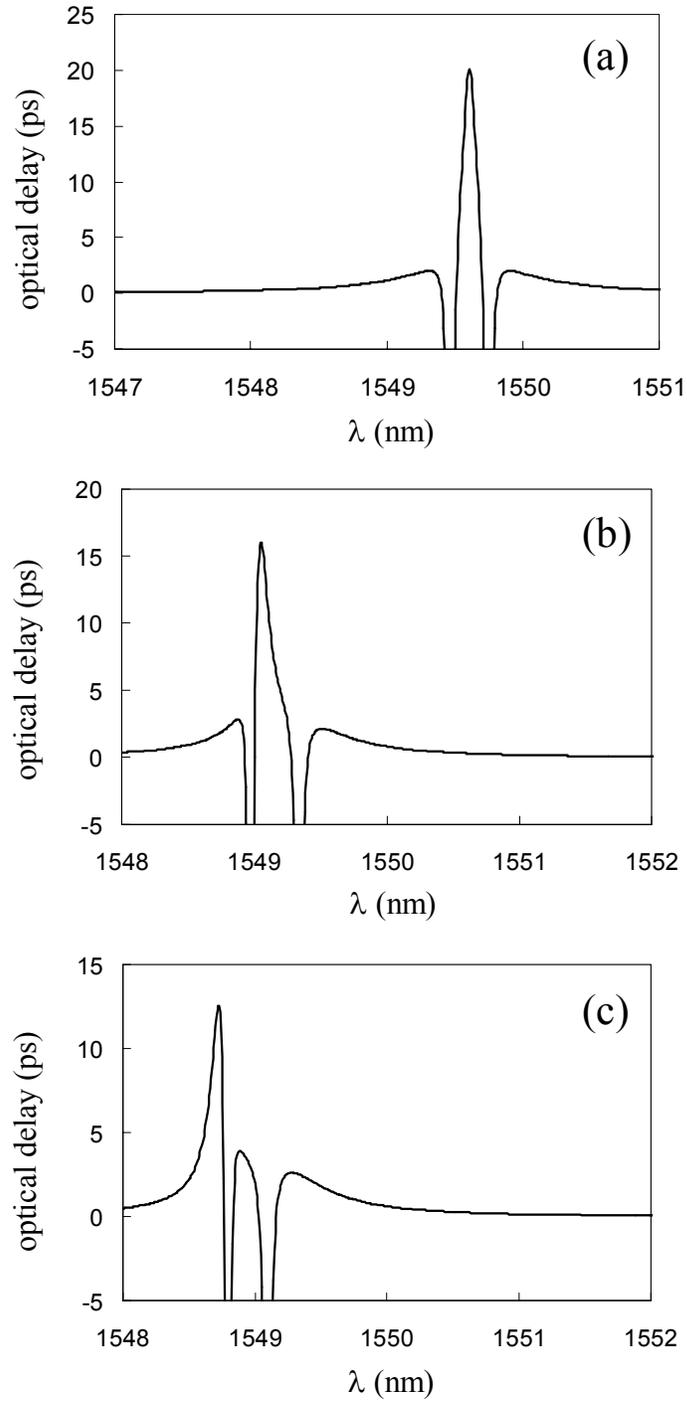